\documentclass{appolb}
\usepackage{graphicx}
\usepackage{amsmath, amsthm}
\usepackage{bm}

\usepackage{ulem}

\renewcommand\sout{\bgroup \color{red} \ULdepth=-.5ex \ULset}

\begin{document}
\title{\bf What makes the peak structure of the $\Lambda p$ invariant-mass
  spectrum in the $K^{-} {}^{3} {\rm He} \to \Lambda p n$ reaction?%
  \thanks{Presented at the {\it 2nd Jagiellonian Symposium on
      Fundamental and Applied Subatomic Physics}, June 4-9, 2017,
    Krak{\'o}w, Poland.}%
}
\author{Takayasu Sekihara
  \address{Advanced Science Research Center, Japan Atomic Energy Agency,
    Shirakata, Tokai, Ibaraki, 319-1195, Japan}
  \\ ~ \\
  Eulogio Oset
  \address{Departamento de F\'{\i}sica Te\'orica and IFIC, Centro
    Mixto Universidad de Valencia-CSIC, Institutos de Investigaci\'on
    de Paterna, Aptdo. 22085, 46071 Valencia, Spain}
  \\ ~ \\
  Angels Ramos
  \address{Departament de F\'{\i}sica Qu\`antica i Astrof\'{\i}sica
    and Institut de Ci\`encies del Cosmos, Universitat de Barcelona,
    Mart\'i i Franqu\`es 1, 08028 Barcelona, Spain}
}
\maketitle
\begin{abstract}
  Recently a peak structure was observed near the $K^{-} p p$
  threshold in the in-flight ${}^{3} {\rm He} (K^{-} , \, \Lambda p)
  n$ reaction of the E15 experiment at J-PARC, which could be a signal
  of a $\bar{K} N N$ bound state.  In order to investigate what is the
  origin of this peak, we calculate the cross section of this
  reaction, in particular based on the scenario that the $\bar{K} N N$
  bound state is indeed generated and decays into $\Lambda p$.  We
  find that the numerical result of the $\Lambda p$ invariant-mass
  spectrum in the $\bar{K} N N$ bound scenario is consistent with the
  J-PARC E15 data.
\end{abstract}

\section{Introduction}

We expect that there should exist kaonic nuclei, i.e., bound states of
antikaon ($\bar{K}$) and usual atomic nuclei, thanks to the strongly
attractive interaction between $\bar{K}$ and nucleon
($N$)~\cite{Kaiser:1995eg, Oset:1997it}.  In particular, the simplest
kaonic nucleus $\bar{K} N N (I = 1/2)$, sometimes called $K^{-} p p$,
has intensively attracted attention in both theoretical and
experimental sides~\cite{Nagae:2016cbm}, but for the moment there has
been no consensus on the properties of the $\bar{K} N N$ bound state.

In this line, the result from the recent J-PARC E15
experiment~\cite{Sada:2016nkb} is very promising.  They observed a
peak structure near the $K^{-} p p$ threshold in the $\Lambda p$
invariant-mass spectrum of the in-flight ${}^{3} {\rm He} (K^{-} , \,
\Lambda p) n$ reaction at the kaon momentum $k_{\rm lab} = 1 \text{
  GeV}/c$.  By fitting this peak with the Breit-Wigner formula, they
reported its mass $M_{X} = 2355 ^{+6}_{-8} \text{(stat.)}  \pm 12
\text{(sys.)} \text{ MeV}$ and width $\Gamma _{X} = 110^{+19}_{-17}
\text{(stat.)} \pm 27 \text{(sys.)} \text{ MeV}$~\cite{Sada:2016nkb}.
This could be a signal of a $\bar{K} N N$ bound state with a binding
$\sim 15 \text{ MeV}$ from the $K^{-} p p$ threshold.

The task in our study is to investigate what is the origin of the peak
in the J-PARC E15 experiment.  In particular, we theoretically check
whether or not the signal of the $\bar{K} N N$ bound state is strong
enough to make a peak structure around the $K^{-} p p$ threshold in
the reaction based on the scenario that the $\bar{K} N N$ bound state
is indeed generated and decays into $\Lambda p$.  In the following, we
calculate the cross section of the $K^{-} {}^{3} {\rm He} \to \Lambda
p n$ reaction at the kaon momentum $k_{\rm lab} = 1 \text{ GeV}/c$.
The details of the calculations are given in
Ref.~\cite{Sekihara:2016vyd}.

\section{The cross section of the $K^{-} {}^{3} {\rm He} \to
  \Lambda p n$ reaction}

\begin{figure}[b]
  \centering
  \begin{minipage}{0.35\hsize}
    \includegraphics[scale=0.18]{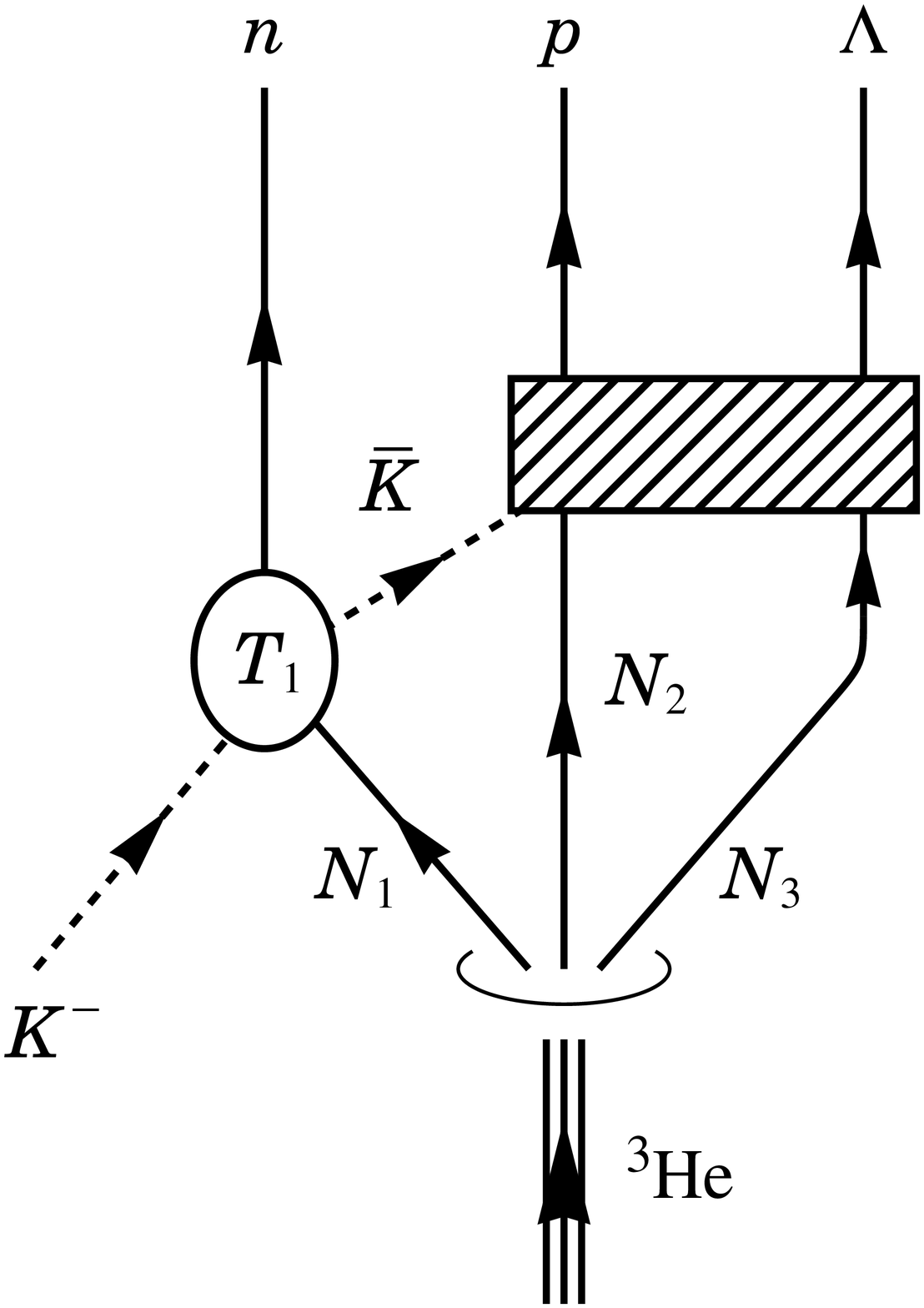}
  \end{minipage}
  \begin{minipage}{0.55\hsize}
    \caption{Feynman diagram most relevant to the three-nucleon
      absorption of a $K^{-}$ in the $\bar{K} N N$ bound
      scenario~\cite{Sekihara:2016vyd}.  We take into account the
      antisymmetrization for the three nucleons in ${}^{3} {\rm He}$.}
    \label{f1}
  \end{minipage}
\end{figure}

\begin{figure}[t]
  \centering
  \includegraphics[width=6.1cm]{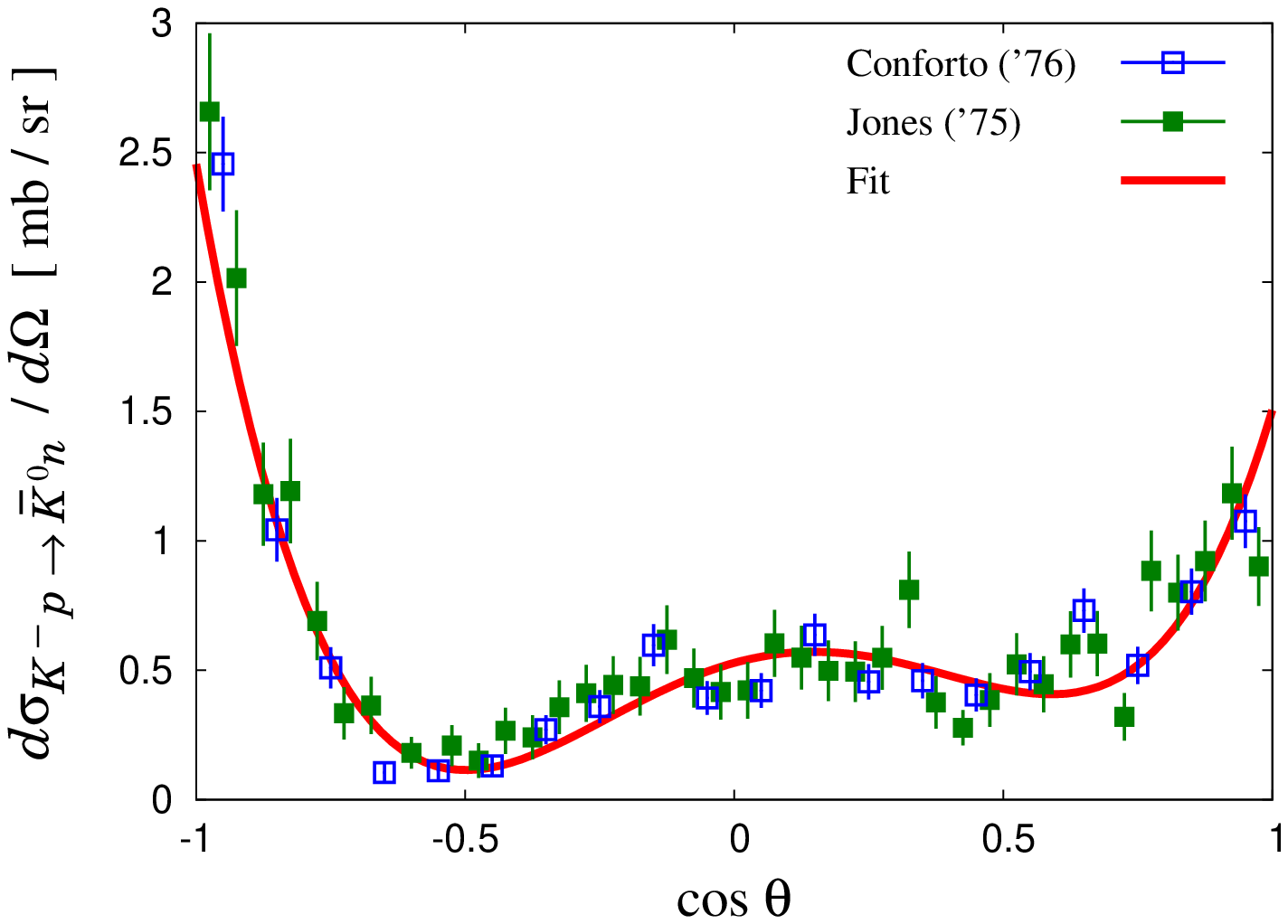} ~
  \includegraphics[width=6.1cm]{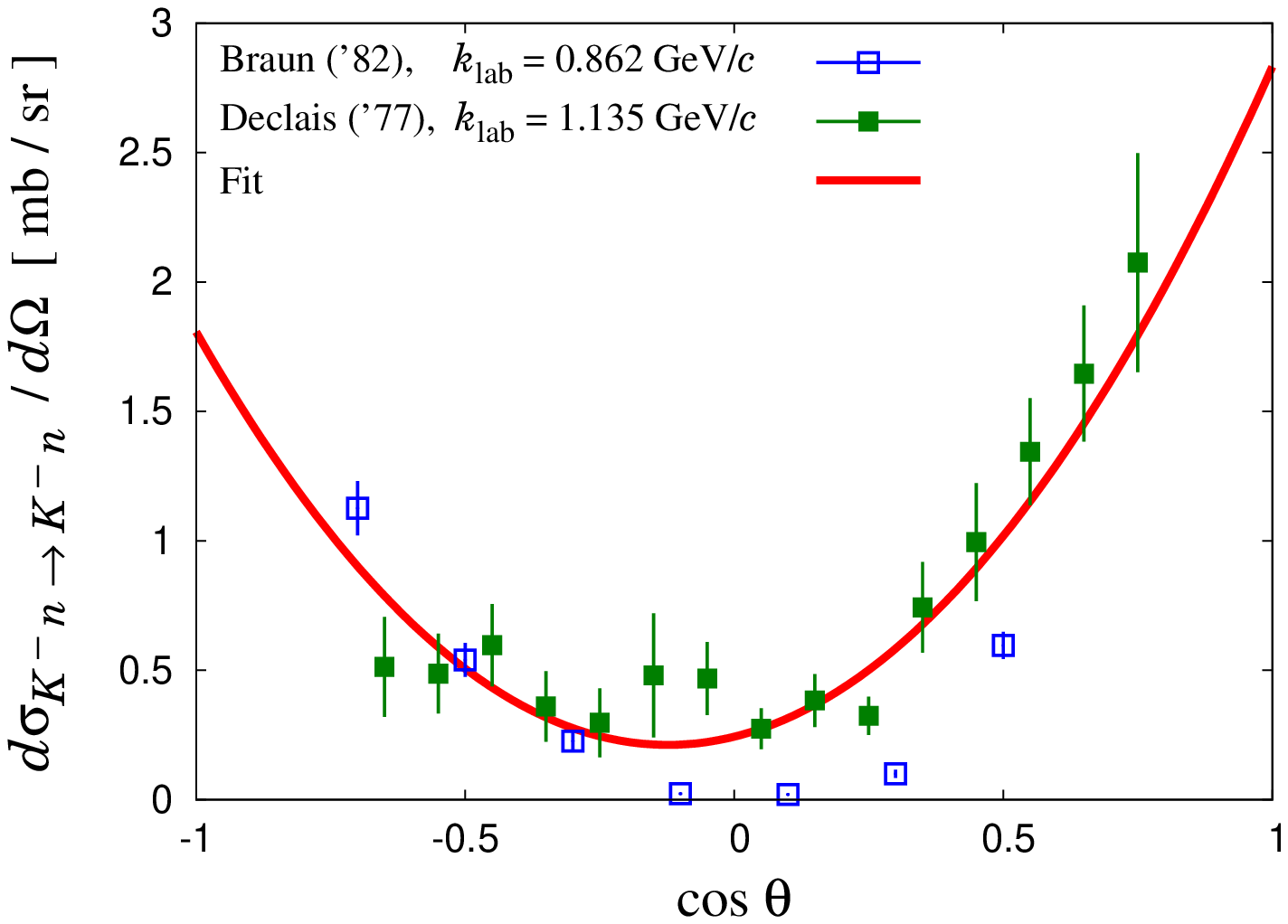}
  \caption{(color online) Differential cross sections of the $K^{-} p
    \to \bar{K}^{0} n$ (left) and $K^{-} n \to K^{-} n$ (right)
    reactions~\cite{Sekihara:2016vyd}, where $\theta$ is the angle of
    the emerging kaon versus the original one.  The experimental data
    are taken from Refs.~\cite{Jones:1974at, Conforto:1975nw} at
    $k_{\rm lab} = 1 \text{ GeV} /c$ for a proton target, and from
    Ref.~\cite{Declais:1977kj} at $k_{\rm lab} = 1.138 \text{ GeV} /c$
    and from~\cite{Braun:1982ih} at $k_{\rm lab} = 0.862 \text{ GeV} /
    c$ for a neutron target.}
  \label{f2}
\end{figure}

Let us first pin down the reaction mechanism for $K^{-} {}^{3} {\rm
  He} \to \Lambda p n$.  Because we are interested in the
three-nucleon absorption of a $K^{-}$ with a final-state energetic
neutron going to its forward direction, as in the J-PARC E15
experiment~\cite{Sada:2016nkb}, we consider the diagram shown in
Fig.~\ref{f1}.  In the first step of the scattering, the initial-state
$K^{-}$ kicks out an energetic neutron in its forward direction and a
slow $\bar{K}$ plus two nucleons remain.  The scattering amplitude for
this first step, $K^{-} p \to \bar{K}^{0} n$ or $K^{-} n \to K^{-} n$,
is fixed so as to reproduce its experimental cross section, as in
Fig.~\ref{f2}.  It is important that the cross sections of these
processes have their local or global minima when the final-state
neutron goes forward, i.e., $\cos \theta = -1$ in Fig.~\ref{f2}.
Thanks to this fact, the $K^{-} {}^{3} {\rm He} \to \Lambda p n$
reaction favors the forward neutron emission compared to the
middle-angle emission.

Next, the slow $\bar{K}$ after the first step propagates and is
absorbed by two nucleons.  This $\bar{K}$ propagator, which is
expressed as $1 / [ ( p_{\text{prop}}^{\mu} )^{2} - m_{K}^{2}]$ with
the propagating $\bar{K}$ momentum $p_{\text{prop}}^{\mu}$, can make a
kinematic peak structure in the cross section, as the propagating
$\bar{K}$ can go almost on its mass shell, $( p_{\text{prop}}^{\mu}
)^{2} \approx m_{K}^{2}$.  In terms of the $\Lambda p$ invariant mass
$M_{\Lambda p}$, this peak appears around
\begin{equation}
  M_{\Lambda p} \approx \sqrt{( p_{\text{prop}}^{0} + 2 m_{N})^{2}
    - \bm{p}_{\text{prop}}^{2} } ,
  \label{eq1}
\end{equation}
where we derived this expression by neglecting the Fermi motion of two
nucleons.  The $\Lambda p$ invariant mass~\eqref{eq1} depends on the
scattering angle of the final-state neutron in the global
center-of-mass frame of the $K^{-} {}^{3} \text{He} \to \Lambda p n$
reaction, $\theta _{n}^{\rm cm}$.  At the forward neutron emission,
$\theta _{n}^{\rm cm} = 0^{\circ}$, $M_{\Lambda p}$ in Eq.~\eqref{eq1}
takes its minimum $\approx 2.40 \text{ GeV}$, and it grows as $\theta
_{n}^{\rm cm}$ becomes larger.  We say that this kinematic peak
structure is due to the quasi-elastic $\bar{K}$ scattering in the
first step, as the propagating $\bar{K}$ can go almost on its mass
shell after the first step.

\begin{figure}[t]
  \centering
  \includegraphics[scale=0.18]{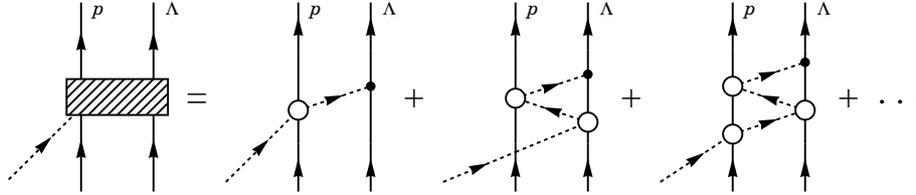}
  \caption{Multiple kaon scattering between two
    nucleons~\cite{Sekihara:2016vyd}, Here dashed lines and open
    circles represent the kaon and the $\bar{K} N \to \bar{K} N$
    amplitude in the chiral unitary approach, respectively.}
  \label{f3}
\end{figure}

Then, let us turn to the absorption process of the slow $\bar{K}$ into
two nucleons, the shaded area in Fig.~\ref{f1}.  In general, we may
consider the multiple scattering of $\bar{K}$ between two nucleons
before the $\bar{K}$ absorption, as shown in Fig.~\ref{f3}.

We note that, even if we take into account only the first term on the
right-hand side of Fig.~\ref{f3}, there is a possibility of making a
peak around the $K^{-} p p$ threshold in the $\Lambda p$
invariant-mass spectrum.  This is because the $\Lambda (1405)$
resonance, which is generated at the open circle in Fig.~\ref{f3} by
the $\bar{K} N$ scattering, exists below the $\bar{K} N$ threshold and
hence the invariant mass of the $\Lambda (1405) p$ system can be lower
than the $K^{-} p p$ threshold.  We call this the uncorrelated
$\Lambda (1405) p$ scenario.  As a result in this scenario, we
obtained the $\Lambda p$ invariant-mass spectrum with a peak at 2370
MeV~\cite{Sekihara:2016vyd}.  Although this peak position is
compatible with the experimental one at $2355_{-8}^{+6} \text{(stat.)}
\pm 12 \text{(sys.)} \text{ MeV}$ within experimental errors, the mass
distribution in the lower energy side clearly falls short of the data.
This indicates the need to improve this scenario by considering new
additional mechanisms that could bring a better agreement with data.

\begin{figure}[t]
  \centering
  \includegraphics[width=8.6cm]{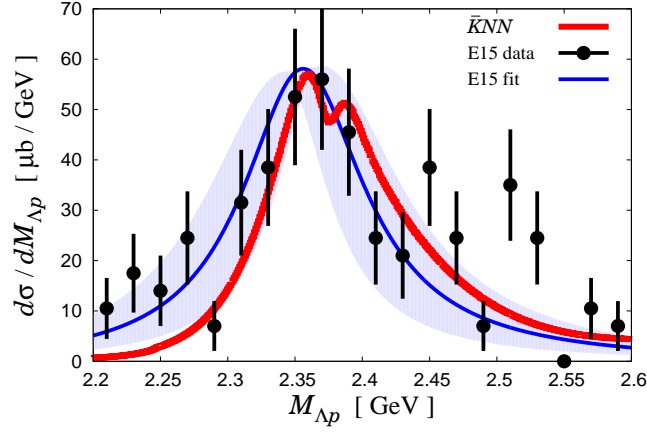}
  \caption{(color online) Mass spectrum for the $\Lambda p$ invariant
    mass of the in-flight ${}^{3}{\rm He} ( K^{-} , \, \Lambda p ) n$
    reaction in the $\bar{K} N N$ bound
    scenario~\cite{Sekihara:2016vyd}.  The experimental (E15) data and
    its fit are taken from Ref.~\cite{Sada:2016nkb} and shown in
    arbitrary units.}
  \label{f4}
\end{figure}

\begin{figure}[b]
  \centering
  \includegraphics[width=8.6cm]{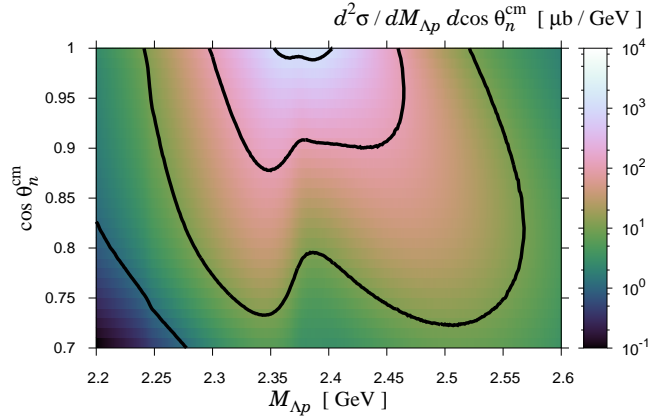}
  \caption{(color online) Differential cross section of the in-flight
    ${}^{3}{\rm He} ( K^{-} , \, \Lambda p ) n$ reaction in the
    $\bar{K} N N$ bound scenario~\cite{Sekihara:2016vyd}.  Contours
    represent $10^{3}$, $10^{2}$, $10^{1}$, and $10^{0}$ $\mu
    \text{b/GeV}$, respectively.}
  \label{f5}
\end{figure}

Now we take into account all the contributions of the summation in
Fig.~\ref{f3}.  Because a $\bar{K} N N$ bound state can be generated
in the full calculation of the multiple scattering in Fig.~\ref{f3},
we call this the $\bar{K} N N$ bound scenario.  In the present
approach, we employ the fixed center approximation to the Faddeev
equation and obtained a resonance pole at $2354 - 36 i \text{ MeV}$ in
the $\bar{K} N N$ scattering amplitude, which corresponds to the
$\bar{K} N N$ bound state.  The resulting $\Lambda p$ invariant-mass
spectrum $d \sigma / d M_{\Lambda p}$ in the $\bar{K} N N$ bound
scenario is plotted in Fig.~\ref{f4} together with the experimental
(E15) data and its fit in arbitrary units~\cite{Sada:2016nkb}.  An
important finding is that our mass spectrum is consistent with the
experimental one within the present error, including the tail at the
lower energy $\sim 2.3 \text{ GeV}$.  We also note that the total
cross section in our calculation $7.6 ~ \mu \text{b}$ is consistent
with the empirical value $7 \pm 1 ~ \mu \text{b}$~\cite{Sada:2016nkb}.

In addition, we observe a two-peak structure below and above the
$K^{-} p p$ threshold in the $\Lambda p$ invariant mass spectrum.  We
have checked that the lower peak at $\sim 2.35 \text{ GeV}$ is the
signal of the $\bar{K} N N$ bound state while the higher peak at $\sim
2.40 \text{ GeV}$ comes from the quasi-elastic kaon scattering in the
first step discussed around Eq.~\eqref{eq1}.  One can see these
origins by plotting the differential cross section $d^{2} \sigma / d
M_{\Lambda p} d \cos \theta _{n}^{\rm cm}$ as shown in Fig.~\ref{f5}.
In this figure, the signal of the $\bar{K} N N$ bound state stays at
$\sim 2.35 \text{ GeV}$ independently of the neutron scattering angle
$\theta _{n}^{\rm cm}$.  On the other hand, the higher peak in
Fig.~\ref{f4} originates from the band which goes from $\sim 2.4
\text{ GeV}$ at $\cos \theta _{n}^{\rm cm} = 1$ to the lower-right
direction in Fig.~\ref{f5}, according to Eq.~\eqref{eq1} as the
contribution from the quasi-elastic scattering of the kaon.

According to all the discussions above, we can say that our results
support the explanation that the E15 signal in the ${}^{3} \text{He} (
K^{-}, \Lambda p ) n$ reaction is likely a signal of the $\bar{K} N N$
bound state.

\section{Summary and outlook}

In this study we have investigated the origin of the peak structure
observed near the $K^{-} p p$ threshold in the ${}^{3} \text{He} (
K^{-}, \, \Lambda p ) n$ reaction of the J-PARC E15 experiment.  The
consideration of the bound $\bar{K} N N$ state brings our results of
the $\Lambda p$ invariant mass spectrum, that contain uncertainties
much smaller than the experimental ones, in consistency with the
experimental band that accounts for the experimental errors, while our
results with the uncorrelated $\Lambda (1405) p$ scenario are clearly
inconsistent with this band.

Finally we emphasize that the high statistics data are coming from the
second run of the J-PARC E15 experiment~\cite{Iwasaki:2017}, in which
they accumulated about 30 times more data than that in the first
run~\cite{Sada:2016nkb}.  With these high statistics data, we will
able to study more things such as the angular dependence of the signal
and the $\Lambda p / \Sigma ^{0} p$ branching ratio so as to conclude
more clearly the existence of the $\bar{K} N N$ bound state.

\end{document}